# Water Reaction Mechanism in Metal Organic Frameworks with Coordinatively Unsaturated Metal Ions: MOF-74

Kui Tan[1], Sebastian Zuluaga[2], Qihan Gong[3], Pieremanuele Canepa[2], Hao Wang[3], Jing Li[3], Yves J. Chabal[1] and Timo Thonhauser[2]

[1]Department of Materials Science & Engineering,
University of Texas at Dallas, Richardson, Texas 75080
[2]Department of Physics, Wake Forest University, Winston-Salem, North Carolina 27109
[3]Department of Chemistry and Chemical Biology, Rutgers University,
Piscataway, New Jersey 08854

**Abstract**

Water dissociation represents one of the most important reactions in catalysis, essential to the surface and nano sciences [e.g., Hass et al., Science, **1998,** 282, 265-268; Brown et al., Science **2001**, 294, 67-69; Bikondoa et al., Nature **2005**, 5, 189-192]. However, the dissociation mechanism on most oxide surfaces is not well understood due to the experimental challenges of preparing surface structures and characterizing reaction pathways. To remedy this problem, we propose the metal organic framework MOF-74 as an ideal model system to study water reactions. Its crystalline structure is well characterized; the metal oxide node mimics surfaces with exposed cations; and it degrades in water. Combining *in situ* IR spectroscopy and first-principles calculations, we explored the MOF-74/water interaction as a function of vapor pressure and temperature. Here, we show that, while adsorption is reversible below the water condensation pressure (~19.7 Torr) at room temperature, a reaction takes place at ~150 °C even at low water vapor pressures. This important finding is unambiguously demonstrated by a clear spectroscopic signature for the direct reaction using $D_2O$, which is not present using $H_2O$ due to strong phonon coupling. Specifically, a sharp absorption band appears at 970 $cm^{-1}$ when $D_2O$ is introduced at above 150 °C, which we attribute to an O-D bending vibration on the phenolate linker. Although $H_2O$ undergoes a similar dissociation reaction, the corresponding O-H mode is too strongly coupled to MOF vibrations to detect. In contrast, the O-D mode falls in the phonon gap of the MOF and remains localized. First-principles calculations not only positively identify the O-D mode at 970 $cm^{-1}$ but derive a pathway and kinetic barrier for the reaction and the final configuration: the D (H) atom is transferred to the oxygen of the linker phenolate group, producing the notable O-D absorption band at 970 $cm^{-1}$, while the OD (or OH) binds to the open metal sites. This finding explains water dissociation in this case and provides insight into the long-lasting question of MOF-74 degradation. Overall, it adds to the understanding of molecular water interaction with cation-exposed surfaces to enable development of more efficient catalysts for water dissociation.

## 1. Introduction

Water dissociation is the key step in many physicochemical phenomena and processes, such as corrosion, passivation, dissolution, precipitation, catalysis, electrochemistry, and environmental chemistry.[1-8] Understanding the interaction or reaction of water with solid materials, such as metal or metal-oxide surfaces, can benefit many fields, especially development of efficient catalysts for water splitting.[1,5,8] Many experimental and theoretical studies have considered the fundamental aspects of water dissociation on oxide surfaces that include $Al_2O_3$,[3] $MgO$,[5,6,9] $ZnO$,[7] and $TiO_2$[1,10] as well as the location of the water molecules, the role of surface defects (oxygen vacancy), and intermolecular interactions. However, detailed understanding has been impeded by the lack of structure/reaction characterization.[4,5,8,11] While water dissociation has been observed on many cation-terminated surfaces or surfaces with defects,[3,12-17] dissociation pathways remain an open question. As a result, we have little direct evidence that open metal sites (e.g., exposed cations at oxygen vacancies) are effective in dissociating water molecules.[1,8,10]

Metal organic frameworks (MOFs) are a family of porous materials usually formed by solvothermal reactions between organic and inorganic species that assemble into crystalline networks.[18] These defined, microporous structures (pores < 2 nm) have high surface areas (up to 5900 $m^2$/g) and pore volume (up to 2 $cm^3$/g) and hold great promise for such applications as gas storage, separation, and purification.[18,19] MOF-74 [$M_2$(dobdc), M=$Mg^{2+}$, $Zn^{2+}$, $Ni^{2+}$, $Co^{2+}$, and dobdc=2,5-dihydroxybenzenedicarboxylic acid] is a well-studied framework with a high density of coordinatively unsaturated metal centers (also called open metal sites) in metal-oxide pyramid clusters.[20] These exposed cations provide charged binding sites that enhance the guest-host interaction for small molecules (e.g., $H_2$, $CO_2$, $CH_4$).[21-26] Furthermore, MOF-74 provides a well-characterized model system to study water's interaction with exposed open sites.

MOF-74 compounds were once considered relatively robust framework materials with good stability in water vapor relative to other prototypical MOFs.[27] However, many recent studies indicate that their gas uptake (e.g. $N_2$, $CO_2$) deteriorates significantly upon exposure to water vapor.[28-32] Upon hydration, gas sorption measurements show that MOF-74 compounds lose a substantial fraction of their original gas uptake capacity ($N_2$, $CO_2$) even at 9% relative humidity and with subsequent thermal regeneration to remove adsorbed water molecules. These results point to an irreversible reaction within MOF-74, but X-ray diffraction measurements and infrared spectroscopy have not been able to detect a notable structural change or chemical transformation following hydration and thermal regeneration.[28,31,32] *In situ* molecular-level characterization is needed to uncover how water interacts with the metal centers in MOF-74.

In this work, a combination of *in situ* IR absorption, Raman scattering, and X-ray diffraction measurements did not detect a hydrolysis reaction at room temperature after water exposure below condensation vapor pressure (~19.7 Torr at 298 K). However, using *heavy water* under the same conditions, we *did* observe a spectroscopically distinct infrared absorption band at ~970 $cm^{-1}$. With the aid of first-principles calculations, this discovery led us to the conclusion that water does dissociates at the metal center of MOF-74 at temperatures between 150 and 200 °C. As we will show in great detail through this work, this finding is of tremendous significance for two reasons: i) once the water molecule dissociates at the metal center, the OH species remains bonded to this site, which is thus passivated. This explains the loss in gas uptake capacity after the MOF is exposed to water vapor. And, ii) while H binds in water to the oxygen with 5.11 eV, after the dissociation it binds to the MOF with only 1.01 eV—a huge improvement, opening up exciting opportunities to "harvest" that loosely bound hydrogen for the purpose of hydrogen production. MOF-74 is thus an outstanding catalyst for water splitting, with the onset of the reaction at only 150 ˚C. This important finding is very fortuitous in that the hydrogen dissociation from heavy water has a very distinct signature in the IR spectrum that falls into the phonon gap of the MOF and is thus clearly observable. The reaction with water itself follows of course the same dissociation pathway, but has gone unnoticed until now as its signature is strongly couple to the MOF and thus not detectable.

## 2. Experimental and theoretical methods

***Sample Preparation (MOF-74):*** The synthesis of (Zn, Mg, Ni, Co)-MOF-74 and its physical characterization (X-ray diffraction) are described in the Supporting Information, as they follow standard procedures reported in numerous publications.

***Infrared spectroscopy:*** Powder MOF-74 (~2 mg) was lightly pressed onto a KBr pellet (~1 cm diameter, 1-2 mm thick), and the crystals evenly dispersed (see Figure S2). This procedure ensures that the data are reproducible, as shown in Figure S9 and our previous studies on the adsorption of small gas molecules ($H_2$, $CO_2$, $CH_4$) on MOF materials.[33-36] The pellet was then placed into a high-pressure, high-temperature cell purchased from Specac (product number P/N 5850c, UK) at the focal point of the sample compartment of the infrared spectrometer (Nicolet 6700, Thermo Scientific, US) equipped with a liquid $N_2$-cooled MCT-B detector. The cell was connected to a vacuum line (base pressure ~10-20 mTorr). The samples were activated by evacuation at 180 ˚C for at least 3 h, until IR measurements showed that the $H_2O$ pre-adsorbed during sample preparation was fully removed. Water vapor was then introduced at pressures varying from 500 mtorr to 8 Torr (40%, RH) at room temperature, which is sufficient to saturate all of the open metal sites of an empty MOF-74.[37] The exposure pressure at a pressure below 8 Torr avoids water condensation onto the cell window and KBr substrate. For temperature dependence studies, the pressure was set at 8 Torr at 100 ˚C, 150 ˚C, and 200 ˚C, all for 20 min. All spectra were recorded in transmission mode with a frequency range of 400-4000 $cm^{-1}$ (4 $cm^{-1}$ spectral resolution) and referenced to the initial clean, activated MOF under vacuum at each temperature.

The effect of water reaction on the gas adsorption capacity of MOF-74 samples was studied, first, by loading the sample with 6 Torr $CO_2$, either before or after water ($H_2O$, $D_2O$) exposures at 200 ˚C, a temperature critical to inducing water dissociation. This pressure (6 Torr) was maintained for 30 minutes, and the gas was then pumped out before IR measurements were carried out. Consequently, all $CO_2$ gas was removed, and only the decomposition products could be identified. All IR spectra were referenced to the initial clean, activated MOF-74 sample.

***Powder X-ray Diffraction:*** Powder X-ray diffraction was used to characterize the crystal structure of the MOF samples after exposure to water vapor. Out-of-plane diffraction data were recorded in the 2 theta mode from 3˚ to 40˚ on a Rigaku Ultima III diffractometer (Cu Kα radiation, X-ray wavelength of 1.5418 Å, operating at 40 keV with a cathode current of 44 mA).

The samples were measured by XRD before water vapor exposure for reference. After exposure, the samples were taken out and remeasured to check on their structural integrity.

***Raman Spectroscopy:*** Raman spectroscopy was used to examine the structural evolution of MOF samples following water vapor exposure. The spectra were recorded using a Nicolet Almega XR Dispersive Raman spectrometer from Thermo Fisher Scientific, Inc. A 780-nm laser was used for excitation, and the output power reduced to 1% (0.04 mW for 780 nm excitation) to avoid sample decomposition induced by laser heating.

***First-principles calculations:*** Calculations used density functional theory with the VASP 5.3.3 code.[38,39] We used a plane-wave basis together with projector-augmented wave pseudo potentials[39] and a kinetic-energy cutoff of 600 eV. vdW-DF accounted for important van der Waals interactions,[40-42] which has proven very accurate in several related MOF-74 studies.[43-47] All systems were optimized until the forces acting on each atom were smaller than 1 meV/Å. Due to the large dimensions of the unit cell, including 54 atoms plus the guest molecules, only the Γ point was sampled. The Hessian matrix and the vibration frequencies were calculated using the finite difference method, displacing atoms in each direction by ±0.015 Å. The vibrations of the molecules adsorbed at the metal centers were calculated by freezing the whole system except for the guest molecules. To ensure that this approximation is accurate, calculations also allowed the whole system to vibrate, with no appreciable change in frequency. In cases where the D and H atoms interact with the linker, all atoms in the linker were free to vibrate, while the rest of the MOF was kept frozen. Vibrational frequency analysis was performed using J-ICE software.[48]

## 3. Experimental Results

Infrared spectroscopy complements isotherm, diffraction, and GC mass spectrometry measurements because it is sensitive to the local bonding inside materials, making it most appropriate to investigate chemical reactions in MOFs. However, detecting water reactions in MOF-74 has been particularly challenging because no vibrational probe (Raman or IR absorption) has been able to detect anything related to water products or framework degradation/protonation under water condensation conditions.

This work made the important discovery that, using heavy water, clear deuteration can be established under specific conditions because of the fortuitous vibrational decoupling described in the following sections. The spectroscopic analysis detected and described subtle shifts in the MOF phonon spectrum itself, as opposed to features related to water or water reaction products, an important distinction. Since MOF-74's stability in water vapor at room temperature has been questioned, we start by unambiguously establishing that there is no hydrolysis reaction at room temperature (Section 3.1), before presenting clear evidence for reaction at or above 150 °C, our main finding (Section 3.2).

## 3.1. Reversible $H_2O$ adsorption in Mg-MOF-74 at room temperature

To study water adsorption in MOF-74 and to determine the spectroscopic signatures upon hydration, the activated Mg-MOF-74 sample was exposed to water vapor as a function of vapor pressure from 50 mTorr to 8 Torr. Figure 1 shows that at the lowest water pressure (50 mTorr), two sharp modes are present at 3663 cm$^{-1}$ and 3576 cm$^{-1}$. We attribute them to the stretching mode of free OH (modes above 3500 cm$^{-1}$) with negligible H-bonding.[49] In this case, the frequencies at ~3663 cm$^{-1}$ and ~3576 cm$^{-1}$ can be ascribed to the asymmetric ($v_{as}$) and symmetric ($v_s$) stretching modes of water molecules coordinated to the metal sites through their oxygen atoms. These findings unambiguously confirm that water adsorbs preferentially at the metal sites, as predicted by theory[44] and shown by diffraction measurements.[50]

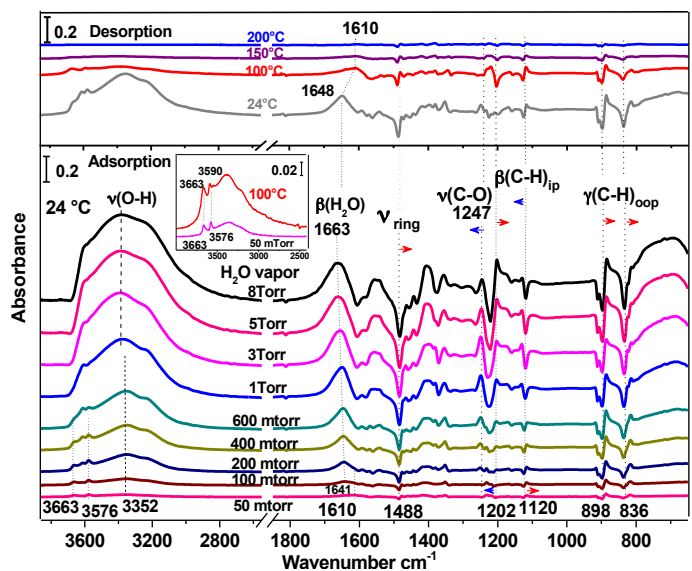

**Figure 1**. IR absorption spectra of $H_2O$ adsorbed into Mg-MOF-74 as function of water vapor pressure (bottom) and upon evacuation (20 mTorr base pressure) as a function of annealing temperature (24 °C, 100 °C, 150 °C, and 200 °C). Each temperature point was held for 20 min. The inset shows the water absorption band at 50 mTorr and annealing below 100 °C. The horizontal arrows indicate the shift (blue, red) of the phonon modes.

Upon increasing the water vapor pressure, these sharp modes gradually disappear and a broad band appears and grows around 3352 cm$^{-1}$, which is associated with hydrogen-bonded water molecules adsorbed in the MOF-74 channels.[49] These additional water molecules are hydrogen bonded to the water molecules adsorbed at the open metal sites, thus broadening and red-shifting the initial sharp modes, $v_{as}$ and $v_s$. In effect, the water molecules that first bond at the open metal sites seed water clustering in the channel. Figure 1 shows that the bending mode of the water molecules, [β($H_2O$)], first appears at 1610 cm$^{-1}$, corresponding to a 15 cm$^{-1}$ blue shift from that of a free water molecule (1595 cm$^{-1}$).[49] As loading increases, the β($H_2O$) mode gradually shifts to 1663 cm$^{-1}$ at 8 Torr, consistent with an increase in hydrogen bonding.[51] The spectral region from 800 to 1600 cm$^{-1}$ is also characterized by strong perturbations of the MOF's phonon modes due to water adsorption.

Table 1 summarizes the induced frequency shifts caused by water adsorption. The red and blue arrows marked in the spectra of Figure 1 correspond to the red and blue shifts listed in Table 1. The frequency positions indicated in the spectra of Figure 1 highlight the shifts; in general, they are not the original positions of the MOF phonon modes as summarized in Table S1 and Table 1. Note that the MOF phonon modes were assigned by comparison with the spectra of pure dobdc ligands and reported salicylate compounds (see Table S1).[52-57]

**Table 1.** Summary of Mg-MOF-74 phonon mode changes after hydration at room temperature.

| MOF phonon mode | Original Position $cm^{-1}$ | Frequency shift ($cm^{-1}$) Low Pressure (i.e. <200 mtorr) | High Pressure (i.e. >5 torr) |
|---|---|---|---|
| $\nu_{ring}(CC)$ | 1484 | -4 | -4 |
| $\nu(C-O)$ | 1238 | -2 | +4 |
| $\beta(C-H)ip$ | 1211 | +2 | -6 |
| $\beta(C-H)ip$ | 1123 | -2 | +3 |
| $\gamma(C-H)oop$ | 895 | -7 | -7 |
| $\gamma(C-H)oop$ | 829 | -4 | -4 |

With slow annealing (6 °C/min) under vacuum, the intensity of the ν(OH) modes gradually decreases, suggesting that the adsorbed water molecules are removed from the frameworks. When they are completely removed, the perturbations of the MOF phonon modes also disappear (see Figure S3). These observations confirm that water is reversibly adsorbed, consistent with Decoste's observation[31] that no protonation reaction product is observed for the carboxylate and phenolate groups after water exposure at room temperature ($H_2O$ in Figure 1 and $D_2O$ in Figure S4). Identical conclusions are drawn for heavy water adsorption at room temperature and subsequent annealing in vacuum (see supporting information).

## 3.2. $H_2O$ and $D_2O$ adsorption in MOF-74 at 100 °C, 150 °C, and 200 °C: Evidence for reaction above 150 °C

The interaction between the water and the MOF-74 compounds at higher temperatures was examined. The samples were annealed at 100 °C, 150 °C, and 200 °C; then 8 Torr of water vapor was introduced in the cell for 20 min before evacuation to ~20 mTorr. The spectra were recorded both during the water vapor loading and immediately after evacuation of gas phase (<2 min).

Figure 2 shows the IR absorption spectra of Mg-MOF-74 after exposure to water and heavy water vapor at different temperatures for 20 min, measured at certain temperatures (100 °C, 150 °C and 200 °C). We see that water molecules are weakly adsorbed at 100 °C, giving rise to absorption bands at 3668, 3586, 3380, and 3215 $cm^{-1}$. The region below 1600 $cm^{-1}$ is dominated by strong perturbations of the MOF-74 vibrational modes observed at room temperature, consistent with water adsorption (see Table 2). The *in situ* X-ray diffraction data show that, once the structure incorporates water,[50] the framework relaxes slightly. Specifically, the lengths of M-O bonds, including M-O (carboxylate) and M-O (hydroxo), increase by approximately 0.2Å. Thus, the structural modification induced by water adsorption is directly responsible for the observed perturbation of several MOF phonon modes, such as (COO⁻) and ν(C-O). These modes are related to the connecting nodes (phenolate and carboxylate groups).

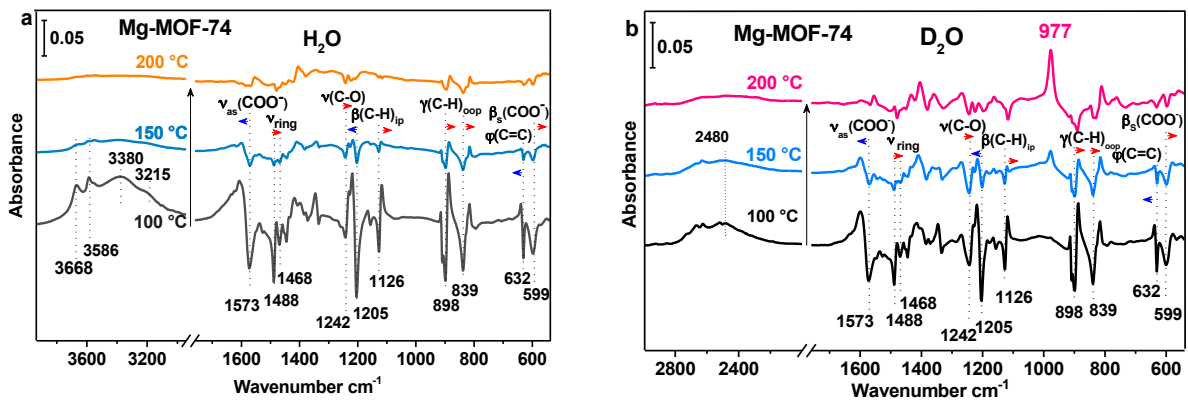

**Figure 2**. IR absorption spectra of hydrated Mg-MOF-74, referenced to the activated MOF in vacuum after introduction of 8 Torr (a) $H_2O$ and (b) $D_2O$ vapor for 20 min and evacuation of gas phase at 100 °C, 150 °C, and 200 °C.

**Table 2.** Summary of changes of Mg-MOF-74 phonon modes upon hydration at 100 °C.

| MOF phonon mode | Original Position cm$^{-1}$ | Frequency shift cm$^{-1}$ |
|---|---|---|
| $\nu_{as}$(COO) | 1580 | +6 |
| $\nu_{ring}$(CC) | 1484 | -4 |
| ν(C-O) | 1238 | -2 |
| β(C-H)ip | 1211 | +3 |
| β(C-H)ip | 1123 | -2 |
| γ(C-H)oop | 895 | -7 |
| γ(C-H)oop | 829 | -7 |
| φ(C=C) | 632 | +2 |
| $\beta_s$(COO⁻) | 587 | -2 |

Similar perturbations of the Mg-MOF-74 phonon modes are observed upon $D_2O$ exposure, consistent with similar adsorption levels (Figure 2b). However, the same figure shows a new distinct vibrational feature at 977 cm$^{-1}$ at 150 °C, which is not present in any part of the spectrum after $H_2O$ loading. Furthermore, as the temperature rises, the peak increases, while the perturbation of the MOF phonons deceases due to the lower density of the adsorbed water molecules.

Similar experiments were repeated for Zn-, Co-, and Ni-MOF-74. For Zn-MOF-74, Figure 3 summarizes the IR absorption spectra: the lower panel for $H_2^{16}O$ and $H_2^{18}O$ exposures at 200°C (see also Figure S5); the upper panel for $D_2O$ exposure at 100 °C, 150 °C, and 200 °C. The sharp band (at 977 cm$^{-1}$ in Mg-MOF-74) is stronger in Zn-MOF-74 upon exposure to $D_2O$ vapor at 200 °C and appears at a slightly lower frequency (970 cm$^{-1}$). At 200 °C, the spectral region above 2000 cm$^{-1}$ shows very little adsorption related to water molecules, suggesting that the mode at 970 cm$^{-1}$ is not associated with an adsorbed water molecule. There are other weak features at 1158, 758, and 621 cm$^{-1}$ in addition to the peak at 970 cm$^{-1}$. The absence of the peak at 970 cm$^{-1}$ when the MOF is exposed to $H_2^{16}O$ and $H_2^{18}O$ confirms that this mode is clearly linked to a deuterium vibration. Similar observations were made for Co- and Ni-MOF-74 (see Figures S6 and S7). In summary, the same peak located at ~970(7) cm$^{-1}$ is observed in all four MOFs samples and is strongest in Zn-MOF-74, followed by Mg-, Ni-, and Co-MOF74, in that order.

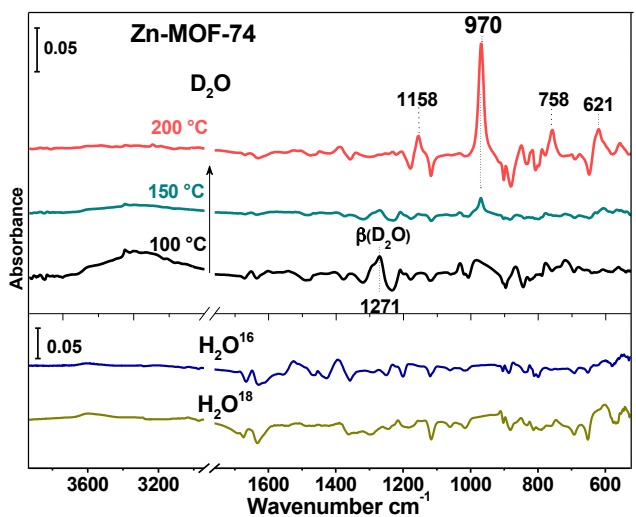

**Figure 3**. (top) IR absorption spectra of Zn-MOF-74 hydrated by introducing 8 Torr $D_2O$ vapor for 20 min and evacuating the gas phase at 100 °C, 150 °C, and 200 °C, referenced to the activated MOF in vacuum. (bottom) Spectra of Zn-MOF-74 exposed to 8 Torr $H_2^{16}O$ and $H_2^{18}O$ under the same conditions at 200 °C, referenced to the activated MOF in vacuum.

To examine the crystal structure of the MOF after exposure to heavy water at different temperatures, X-ray diffraction patterns of the samples were collected before and after. Figure 4 summarizes XRD results for Zn-MOF-74, showing that the major peaks remain unaffected. The strongest peaks at 6.7° and 11.7°, associated with reflections from the $(2\bar{1}0)$ and (300) planes of the original MOF,[50] do not shift (< 0.1°) or broaden (Δ< 0.02°). Hence, the initial crystallinity is maintained after $D_2O$ exposure at high temperature (200 °C). Raman spectra are also useful because they are dominated by the MOF's skeleton phonon modes of the carboxylate, phenolate, and aromatic rings, such as $\nu_{as,s}(COO^-)$, $\beta_{as,s}(COO^-)$ $\nu(C-O)$, and $\nu(C=C)_{aromatic\ ring}$. The carboxylate and phenolate groups are coordinated with metal ions, forming metal-oxygen bonds to support the whole MOF structure. Their vibrational modes are expected to be sensitive to any MOF structural modification. However, the Raman spectra in Figure 5 show no measurable changes (i.e., shift < 0.8 cm$^{-1}$) in the pristine Zn-MOF-74 spectra upon D$_2$O exposure at 200 ˚C, confirming that the MOF crystalline structure is maintained.

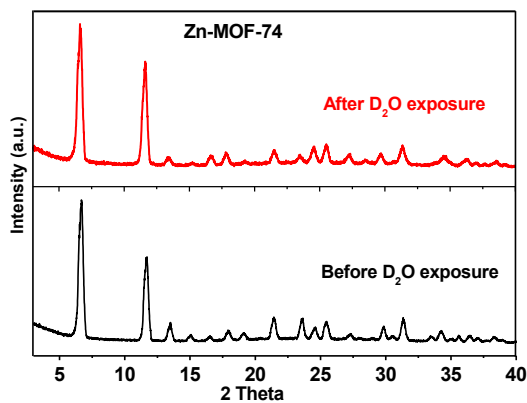

**Figure 4.** Powder X-ray diffraction patterns of activated (bottom) and hydrated (top) Zn-MOF-74 after exposure to 8 Torr D$_2$O vapor at 200 ˚C and cooling to room temperature. The activated (pristine) sample was obtained by annealing to 180 ˚C in a vacuum (< 20 mTorr) for 3 h before cooling to room temperature.

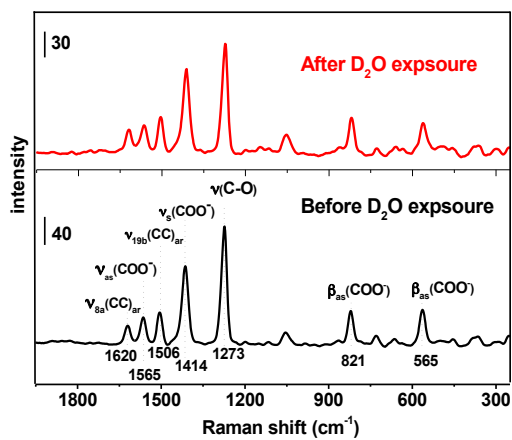

**Figure 5.** Raman spectra of activated (bottom) and hydrated (top) MOF samples after exposure to 8 Torr D$_2$O vapor at 200 ˚C. The spectra were recorded at room temperature.

Next, the MOF gas adsorption capacity was examined after D$_2$O or H$_2$O exposure at 200 ˚C, using CO$_2$ as the probe molecule. The CO$_2$ concentration was measured by *in situ* infrared spectroscopy using a 6 Torr CO$_2$ exposure in Zn-MOF-74 samples with two different pretreatments. The first sample was a fully activated Zn-MOF-74 subjected to slow annealing under vacuum prior to water exposure, as typically for a virgin sample. The second Zn-MOF-74 sample was first exposed to 8 Torr H$_2$O or D$_2$O for 20 min at 200 ˚C, then cooled to room temperature. As shown in Figure 6, CO$_2$ absorbance decreases by ~60% of the initial value in a virgin activated sample pre-exposed to either H$_2$O or D$_2$O. This observation clearly suggests that pre-adsorption of either H$_2$O or D$_2$O leads to similar reduction in CO$_2$ uptake, confirming similar chemistry. However, a sharp vibrational mode can be detected (970 cm$^{-1}$) only after D$_2$O pre-exposure, not after H$_2$O. The origin and significance of this mode is discussed in Section 4.1.

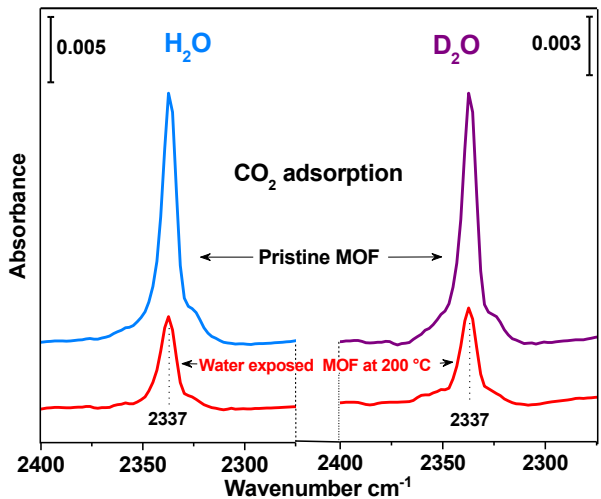

**Figure 6**. IR spectra of CO$_2$ adsorption in activated Zn-MOF-74 (pristine) and exposed to H$_2$O (left) and D$_2$O (right) at 200 ˚C for 20 min. All the measurements were taken after the sample was cooled to room temperature.

For completeness, an estimate of the reactivity of water molecules within the four compounds is derived by normalizing the intensities of this special mode in Figures 3, 7, and S7 to those of the MOF β(C-H) mode around 1200 cm$^{-1}$ (Co-MOF-74, ~0.011; Ni-MOF-74, ~0.027; Mg-MOF-74, ~0.063; Zn-MOF-74, ~0.088), yielding a relative comparison of MOF reactivities: Co-MOF-74 < Ni-MOF-74 < Mg-MOF-74 < Zn-MOF-74. This trend is confirmed by the spectroscopic measurements of CO$_2$ adsorption in pristine samples and samples exposed to D$_2$O at 200 ˚C (see Figures 6 and S8), showing that Co-MOF-74 is the least affected by, or most resistant to, moisture, and Zn-MOF-74 is most reactive. In the case of Zn-MOF-74, even exposure to very low humidity levels (3%, 600 mtorr D$_2$O) at 200 ˚C for 20 min promotes visible growth of the 970 cm$^{-1}$ band, while the CO$_2$ adsorption decreases by ~13% under 6 Torr gas phase (see Figure S10). These conclusions about water reactivity in MOF-74 are generally consistent with Kizze's findings:[28] Mg<Zn<Ni<Co-MOF-74, except for Zn-MOF-74 and Mg-MOF-74, which are reversed, but close.

Finally, for Mg-MOF-74, the 977 cm$^{-1}$ band grew during 4 cyclic D$_2$O exposures at 9.5 Torr and 200 ˚C (see Figure S11) and CO$_2$ adsorption, measured after cooling to room temperature, diminished proportionally. These results indicate that a product of water dissociation is saturating the metal centers and blocking adsorption of other molecules. However, the Raman spectra in Figure 5 and S12 show that none of the crystalline structures are degraded by interaction with D$_2$O at 200 ˚C. The difference spectra in Figure 2 and 3 also show no protonation reaction of carboxylate and/or phenolate groups; for example, formation of ν(C=O) above 1650 cm$^{-1}$ from carboxylic acid groups or ν(C-OD) in the lower frequency region (usually -30 cm$^{-1}$ compared to ν(C-OM)) due to formation of phenolic acid groups.[55-57]

## 4. Discussion

### 4. 1.  Origin of 970 cm$^{-1}$ band

The appearance of a vibrational feature (~970 cm$^{-1}$), which is not present in the regular MOF-74 phonon modes, indicates that a reaction takes place when $D_2O$ is introduced at temperatures above 150 °C in all MOF-74 compounds. The degree of reaction between $D_2O$ and MOF-74 varies with the specific central metal ions as follows: Co-MOF-74 < Ni-MOF-74 < Mg-MOF-74 < Zn-MOF-74, suggesting that the metal ions are involved in the reaction. To confirm this point, the spectrum of the pure dobdc ligand was also examined after $D_2O$ exposure at 200 °C by mixing it in KBr powder and forming a pellet. No mode at ~970 cm$^{-1}$ was detected (Figure S13). Therefore, this mode cannot originate from direct reaction with a linker or the KBr substrate.

The possibility that this mode is associated with adsorption at defect sites (e.g., the external surfaces of crystallites) can also be ruled out for two reasons. First, its intensity (Figure 7) is two orders of magnitude stronger than expected for H at internal surfaces. Second, the adsorption capacity for $CO_2$ in the MOF after $D_2O$ exposure at 200 °C (monitored by its asymmetric stretching mode, $ν_3$) decreases steeply (~60%), which would not be the case if only the surfaces were reacted. The latter observation further confirms that the reaction occurs at the metal center (see Figures 6, S8, S10, S11) since $CO_2$ is known to adsorb at this site.[21,23]

The IR spectroscopic results indicate that this band is only associated with the motion of D atoms. The absence of a band associated with H motion is surprising since the physical and chemical properties of $H_2O$ and $D_2O$ appear identical.[58] Furthermore, the $CO_2$ loading results measured by *in situ* IR (Figure 6) confirm similar reactivity of $H_2O$ and $D_2O$ in MOF-74.

The first clue to this mystery appears in the MOF vibrational spectrum shown in Figures 7 and S7: the sharp mode at ~970 cm$^{-1}$ is always located in the phonon gap of the MOF vibrational spectrum. On the other hand, the corresponding mode for hydrogen is expected to be ~1333 cm$^{-1}$ since the vibrational frequencies in the simple harmonic oscillator approximation are proportional to $\frac{1}{\sqrt{\mu}}$ (where $\mu$ is the reduced mass). This peak would then be located in a spectral region with strong MOF modes; that is, outside the vibrational gap region. This discrepancy opens the possibility of vibrational coupling, typically leading to intensity exchange and broadening.

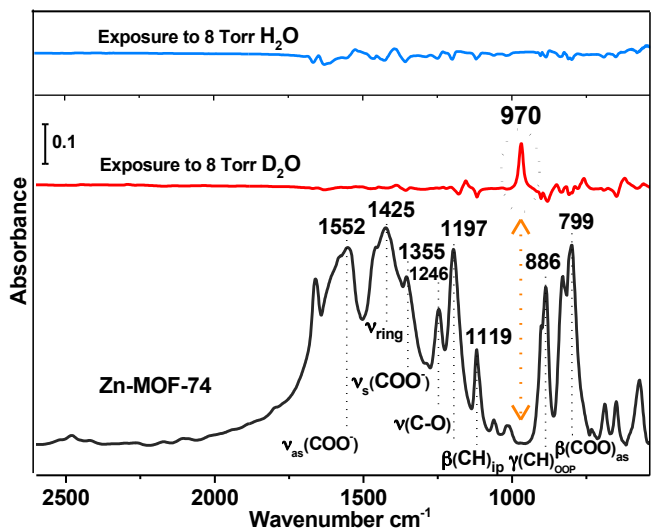

**Figure 7**. IR spectra of Zn-MOF-74 exposed to $H_2O$ (blue) and $D_2O$ (red), referenced to the activated MOF spectra, compared with IR absorption spectra of activated MOF-74, referenced to the KBr pellet.

To determine the origin of the peak and to explain why a similar peak is not present when the experiment is repeated with $H_2O$, we turn to first-principles calculations at the DFT level. We focus here on Zn-MOF-74, as it shows the strongest peak experimentally. The vibrational frequencies were initially calculated for molecularly adsorbed water ($H_2O$, $D_2O$) and any dissociated products (OH, OD, H, and D) adsorbed on the metal center. Table 3 gives the results for one guest molecule per unit cell. We find that the experimentally observed peak at 970 cm$^{-1}$ does not correspond to the vibration of water, heavy water, or any of the products bonded to the metal center. On the other hand, we associate the bending mode of the OD group bonded to the metal center at 582 cm$^{-1}$ with the observed peak at 621 cm$^{-1}$ (Figure 3).

A second clue emerges from Figures 2, 3, and S7. In all four MOFs (Zn-, Mg-, Ni-, and Co-MOF-74), the sharp peak is located no more than 7 cm$^{-1}$ from the 970 cm$^{-1}$ mark. Therefore, the vibrational frequency weakly depends on the nature of the metal center. We conclude that while the origin of this band is not a vibration of a species directly bonded to the metal center, it is associated with a species near the metal center and most likely results from a reaction involving the metal center.

**Table 3.** Calculated vibrational frequencies of H$_2$O and D$_2$O at the metal site and OH, OD, H, and D chemisorbed on the metal center.

| Adsorbed molecule | Asymmetric stretch mode (cm$^{-1}$) | Asymmetric stretch mode (cm$^{-1}$) | Stretch mode (cm$^{-1}$) | Bending mode (cm$^{-1}$) |
|---|---|---|---|---|
| H$_2$O | 3743 | 3635 | -- | -- |
| D$_2$O | 2742 | 2617 | -- | -- |
| OH | -- | -- | 3701 | 779 |
| OD | -- | -- | 2692 | 582 |
| H | -- | -- | 1843 | |
| D | -- | -- | 1303 | |

Based on these observations, we examined the interaction of H and D atoms with the linker. Three stable, nonequivalent configurations for the adsorption of H and D on the linker of the Zn-MOF-74 system were found after the water dissociation reaction, labeled as (1), (2) and (3) in Figure 8. Their vibrational frequencies were calculated for three cases: (i) MOF with no guest molecules, termed *linker*; (ii) MOF with an H atom adsorbed on one of the oxygen atoms of the linker next to the metal center on which an OH molecule is adsorbed, termed *linker+H*; and (iii) analogous to (ii), but with D adsorbed instead of H, termed *linker+D*.

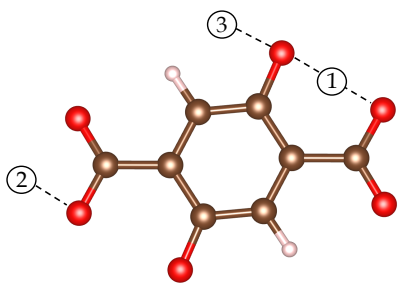

**Figure 8**. MOF-74 linker with three nonequivalent adsorption sites for H and D. Brown = carbon; white = hydrogen; and red = oxygen atoms.

Table 4 lists the vibrational frequencies close to 970 cm$^{-1}$ for *linker+D*. The last column gives the frequencies of *linker* with the same eigenvectors. We find that the two vibrational modes observed at 999 and 1008 cm$^{-1}$ in *linker+D* have corresponding modes in *linker*, both with frequencies at 1029 cm$^{-1}$ (see Figure S15). Therefore, adding D in positions 1 and 2 should decrease the peak at 1029 cm$^{-1}$ and increase the peak at 999 and 1008 cm$^{-1}$. No such differential feature is observed in Figures 3 and 7. Instead, a single band is observed at 970 cm$^{-1}$, away from the ~1029 cm$^{-1}$ region. For position 3, however, Table 4 shows a vibrational mode at 950 cm$^{-1}$ in *linker+D* with no corresponding mode in *linker*. For this vibrational mode, the motion of the D atom dominates, with negligible motion of the linker atoms (Figure 9a). Furthermore, this mode occurs in the middle of the ~100 cm$^{-1}$ gap. This result not only shows a vibrational frequency that matches (within 2%) the experimentally observed peak, but also that, in this case, the vibration of the D atom is isolated (local mode); that is, with negligible coupling to linker vibrations. It explains the sharp adsorption band experimentally observed at 970 cm$^{-1}$ in the Zn-MOF-74 system and confirms the dissociation of the D$_2$O molecule with one D atom attached to the linker's O atom and the remaining OD attached to the uncoordinated metal center.

**Table 4.** Calculated vibrational frequencies of the systems *linker+D* and *linker*. The adsorption sites are labeled according to Figure 8.

| Adsorption site | Frequency of *linker +D* (cm$^{-1}$) | Corresponding frequency of the system *linker* |
|---|---|---|
| 1 | 999 | 1029 |
| 2 | 1008 | 1029 |
| 3 | 950 | Missing |

Now we address why the mode corresponding to 970 cm$^{-1}$ cannot be detected easily in Figure 7, when H$_2$O is introduced instead of D$_2$O. The isotopically shifted mode for *linker+H* is expected around 1333 cm$^{-1}$. In fact, a quick search reveals four vibrational frequencies at 1371, 1332, 1322 and 1317 cm$^{-1}$ (Figure S16). However, a symmetry analysis of the eigenvector reveals that none corresponds to the vibrationally pure mode observed in *linker+D* at 950 cm$^{-1}$; all four involve significant movement of all atoms (C and H) in the linker. The mode at 1317 cm$^{-1}$ most resembles *linker+D* at 950 cm$^{-1}$; it is the one with a strong H-bending motion. However, this *linker+H* mode involves significant motion of the C atoms. Note that the pristine MOF has a degenerate mode with the same frequency (1317 cm$^{-1}$) and a different eigenvector (Figure 9b, c; an animation of vibrational modes in video format is also available). In fact, the entire region around 1330 cm$^{-1}$ is highly populated by linker modes. It follows that the mode at 1317 cm$^{-1}$ and the other three vibrational modes for *linker+H* are strongly coupled to the vibrations of the pristine linker and cannot be observed experimentally.

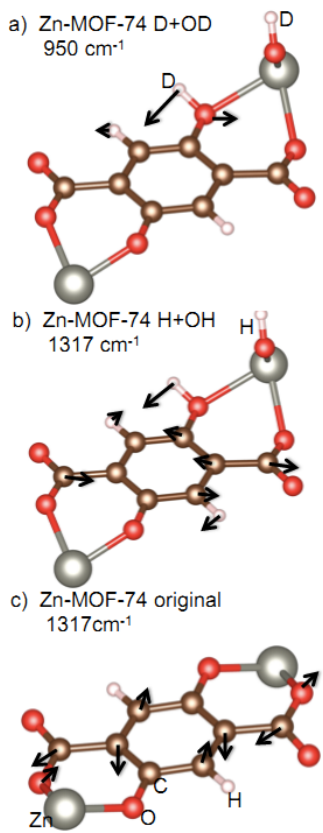

**Figure 9**. Calculated vibrational modes of (a) Zn-MOF-74 with dissociated $D_2O$ (D + OD) at 950 cm$^{-1}$; (b) Zn-MOF-74 with dissociated $H_2O$ (H + OH) at 1317 cm$^{-1}$; and (c) Zn-MOF-74 (original structure) at 1317 cm$^{-1}$. The black arrows represent the eigenvector of the vibrational mode.

We conclude that in the case of $D_2O$, the new vibrational mode of linker+D can be detected because its frequency lies in the gap of the vibrational MOF spectrum. In the case of $H_2O$, the same mode is undetectable because the vibrations of linker+H are coupled to the motion of the carbon atoms in the linker and broadened due to the close proximity of many other modes.

### 4.2. Water dissociation pathway

Henderson's review paper[8] noted that a main requirement for water dissociation on oxide surfaces is a strong bond between the oxygen atom of the $H_2O$ molecule and the cation site and a short distance between the $H_2O$ molecule and the substrate. Under these conditions, the H atoms can be transferred from the water molecule to the surface. Defects, such as oxygen vacancies, are usually considered the most reactive sites for dissociating water molecules since exposed cation sites are highly undercoordinated and energetically favorable for binding water molecules.[1,8] We can quantify this process, defining the exact pathway of water dissociation in MOF-74.

The local chemistry environment of coordinatively unsaturated metal centers in MOF-74 resembles a metal oxide surface with exposed cations. Based on Henderson's suggestion,[8] we optimized the geometry of a single water molecule adsorbed in the $Zn^{2+}$ metal site. Examining the charge density redistribution upon adsorption in Figure 10, we found an excess of charge between the oxygen of the water molecule and the metal center, indicating a covalent bond. We also found that H atoms of the water molecule experienced a loss of charge. Figure 10 also shows a strong charge-density redistribution around the O atoms of the linker, indicating a hydrogen bond between water molecule and linker. We conclude that the heavy-water dissociation mechanism in MOF-74 starts when the $D_2O$ attaches to the metal center. The D atoms then rotate to establish a hydrogen bond with the O of the linker. With an increase in temperature the water molecule dissociates into D and OD. The D atom is picked up by the oxygen of the pheonate group, causing the notable absorption band at 970 $cm^{-1}$, while the OD remains bonded to the metal sites. A weak absorption band at 621 $cm^{-1}$ (see Figure 3 for OD bonded to $Zn^{2+}$), stronger than the O-D stretch band, provides evidence for OD bonding to the metal center.

To estimate the energy barrier and find the transition state for the water dissociation in Zn-MOF-74, we performed climbing-image nudged-elastic band (NEB) calculations and found an energy barrier of 1.01 eV (Figure 11). Note that once the water molecule has been dissociated, the reverse process—transport of the D atom from the linker back to the OH at the metal center (OD+D -> $D_2O$)—has an energy barrier of only 0.025 eV. Of course, the $H_2O$ molecule follows the same dissociation pathway, but our experiments cannot detect the fingerprint for the reasons explained above. Note that the MOF structure is not perturbed by water dissociative adsorption. The CO-M bond is elongated but holds after H or D atoms bind to the linker's oxygen, so the original phonon modes of the MOF are not modified (Figures 3, and 5). Finally, an important consequence of the passivation of the metal sites by the hydroxyl OH (or OD) group it that the gas uptake is strongly affected (Figures 6, S8, and S10, S11).

Changing the metal center should not alter the mechanism but just modulate the reaction barriers slightly since the other Mg-, Co-, Ni-MOF-74 are isostructural with a vibrational band gap in the same region (Figures 7, and S7). All exhibit the same fingerprint mode at ~970 $cm^{-1}$, and a full set of calculations for each was not deemed necessary.

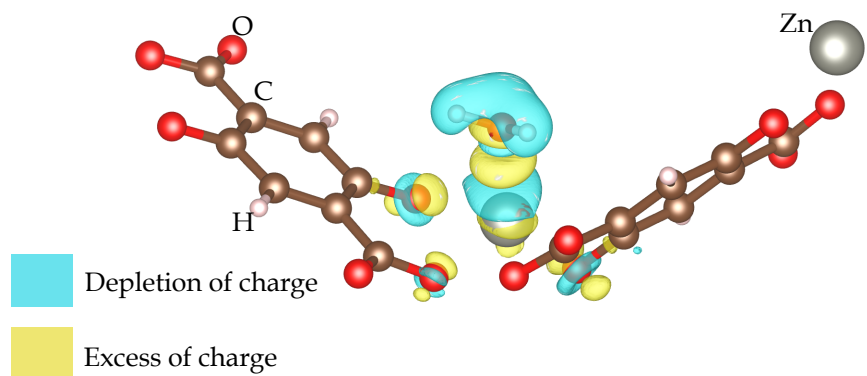

**Figure 10**. Charge density redistribution after water adsorption on Zn-MOF-74. Blue areas denote depletion; yellow areas accumulation of charge. Iso-surfaces were set to 0.01 e/Å$^3$. Some MOF elements were removed for visualization purposes.

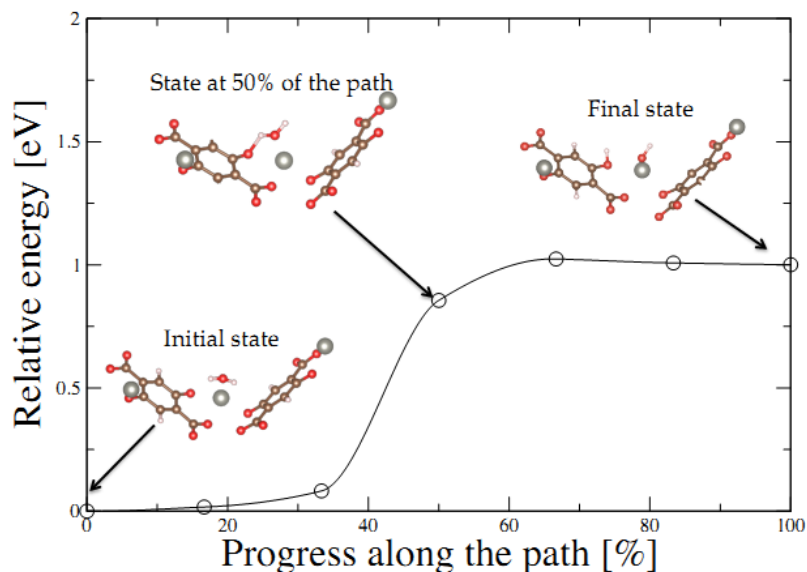

**Figure 11.** Energy barrier for water dissociation on the metal center of Zn-MOF-74. The inserted figures show the initial and final configurations.

In summary, MOF-74 is an ideal platform to test Henderson's model. The structure is well characterized and decorated with open cation sites for binding water molecules. H-bonding between O atoms of the linker and the hydrogen of the adsorbed water is weak, with a distance of 3.16 Å, and the coordinative bond between the O atom and the metal center is strong, constituting a precursor for the water dissociation reaction. The fact that different metal centers have different reactivities (Zn>Mg>Ni>Co) confirms the role of open metal ions in breaking water molecules.

## 5. Conclusions

The reaction of water in MOF-74 has been identified by the observation, using *in situ* infrared spectroscopy, of a sharp band at 970 cm$^{-1}$ after exposure to heavy water ($D_2O$) above 150 °C. DFT calculations show that this new band is associated with the bending vibration of a D atom bonded to one of the O atoms of the linker. The calculations also explain why the corresponding band at ~1300 cm$^{-1}$ is not observed for $H_2O$ adsorption: the vibrational modes of the added H atom are strongly coupled to the vibrations of the linker. The water dissociation mechanism in MOF-74 is based on two elements: (1) the covalent bond between the water molecule and the metal center; and (2) hydrogen bonding between the O atoms of the linker and the H atoms of the water molecule. This work clearly shows that coordinatively unsaturated metal structures in MOF materials are able to break the water molecules at moderate temperatures. The reactivity depends on the type of metal ions: Zn is more reactive than Co. After water dissociation, the open metal ion is occupied by the dissociation products (OH or OD), hindering further adsorption of other molecules, such as $CO_2$, with significant implications for gas sequestration or capture applications. Along the same lines, our results further explain discrepancies reported in the literature—if the sample is annealed too fast or in a poor vacuum, water molecules can be trapped in the crystal when the temperature reaches 150 °C. At that point, they dissociate into OH and H at the metal centers and prevent other molecules from attaching there.

**Supporting information:** Sample preparation; IR and Raman spectra of MOF phonon modes; Raman spectra of hydrated Mg, Ni, and Co-MOF-74; IR absorption spectra of Zn-MOF-74 hydrated by introduction of 8 Torr $D_2O$ vapor at 200 ˚C; $CO_2$ adsorption in MOF-74 before and after $D_2O$ exposure; calculated vibrational modes for original Zn-MOF-74 and Zn-MOF-74 with dissociated $D_2O$ species This material is available free of charge at http://pubs.acs.org.

**Corresponding Author**
Yves J. Chabal: chabal@utdallas.edu  & Timo Thonhauser: thonhauser@wfu.edu

**ACKNOWLEDGMENT**
This work was entirely supported by the Department of Energy Grant No. DE-FG02-08ER46491.

**Table of Contents**

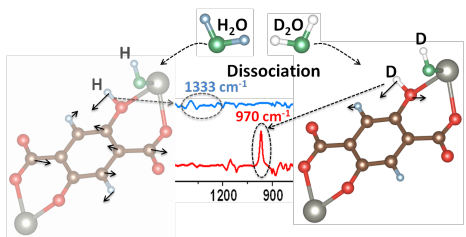

– Supporting information –
# Mechanism of Water Reaction within Metal Organic Frameworks with Coordinatively Unsaturated Metal Ions: MOF-74

Kui Tan[1], Sebastian Zuluaga[2], Qihan Gong[3], Pieremanuele Canepa[2], Hao Wang[3], Jing Li[3], Yves J. Chabal[1] and Timo Thonhauser[2]

[1]Department of Materials Science & Engineering, University of Texas at Dallas, Richardson, Texas 75080
[2]Department of Physics, Wake Forest University, Winston-Salem, North Carolina 27109
[3]Department of Chemistry and Chemical Biology, Rutgers University, Piscataway, New Jersey 08854

## 1. Sample and pellet preparation

Zn-MOF-74: A mixture of zinc nitrate hexahydrate (0.24 g, 0.8 mmol), 2, 5-dihydroxyterephthalic (0.08 g, 0.4 mmol), 9 ml DMF and 1 ml $H_2O$ were transferred into a 28 ml Teflon-lined autoclave. The autoclave was then sealed and heated to 120 ˚C for 3 days. After filtering and washing with 20 ml DMF, the product was collected. Then the product was exchanged with 20 ml methanol in a glass vial every 2 hs during daytime for 3 days.

Mg-MOF-74: A mixture of magnesium nitrate hexahydrate (0.26 g, 1 mmol), 2, 5-dihydroxyterephthalic (0.1 g, 0.5 mmol), 7 ml tetrahydrofuran (THF), 3 ml water $H_2O$ and 2 ml 1 M NaOH solution was prepared in a 28 ml Teflon-lined autoclave. The autoclave was then sealed and heated to 100 ˚C for 3 days. After filtering and washing with 20 ml THF, the product was collected and exchanged with methanol every 2 hs during daytime for 3 days. Then the MOFs sample was stored in $N_2$ glove box.

Ni-MOF-74: A mixture of nickel nitrate hexahydrate (0.24 g, 0.8 mmol), 2, 5-dihydroxyterephthalic (0.08 g, 0.4 mmol), 9 ml DMF and 1 ml $H_2O$ was prepared in a 28 ml Teflon-lined autoclave. The autoclave was then sealed and heated to 100 ˚C for 3 days. After filtering and washing with 20 ml DMF, the product was collected and exchanged with methanol every 2 hs during daytime for 3 days. Then the MOFs sample was stored in $N_2$ glove box.

Co-MOF-74: A mixture of cobalt nitrate hexahydrate (0.17 g, 0.6 mmol), 2, 5-dihydroxyterephthalic (0.06 g, 0.3 mmol), 9 ml DMF and 1 ml $H_2O$ was prepared in a 28 ml Teflon-lined autoclave. The autoclave was then sealed and heated to 100 ˚C for 3 days. After filtering and washing with 20 ml DMF, the product was collected and exchanged with methanol every 2 hs during daytime for 3 days. Then the MOFs sample was stored in $N_2$ glove box.

The crystal structures of the four MOF-74 samples (Zn, Mg, Co, Ni) were measured by PXRD and compared to the simulated PXRD pattern of Ni-MOF-74 in reference 1, as shown in Figure S1a and Figure 4. The XRD diffraction patterns of the samples we studied are in agreement with literature reports. The broadened peaks in Ni, Co-MOF-74 indicate smaller crystal sizes in these two powder samples than in Mg-MOF-74. The BET surface areas of the four MOF-74 samples (Zn, Mg, Co, Ni) were measured (see Figure 1b) and the values for Mg-, Zn-, Co-, and Ni-MOF-

74 are 1078, 774, 1077, and 913 m²/g (see the table below), all within the range reported in the literature.[1-8]

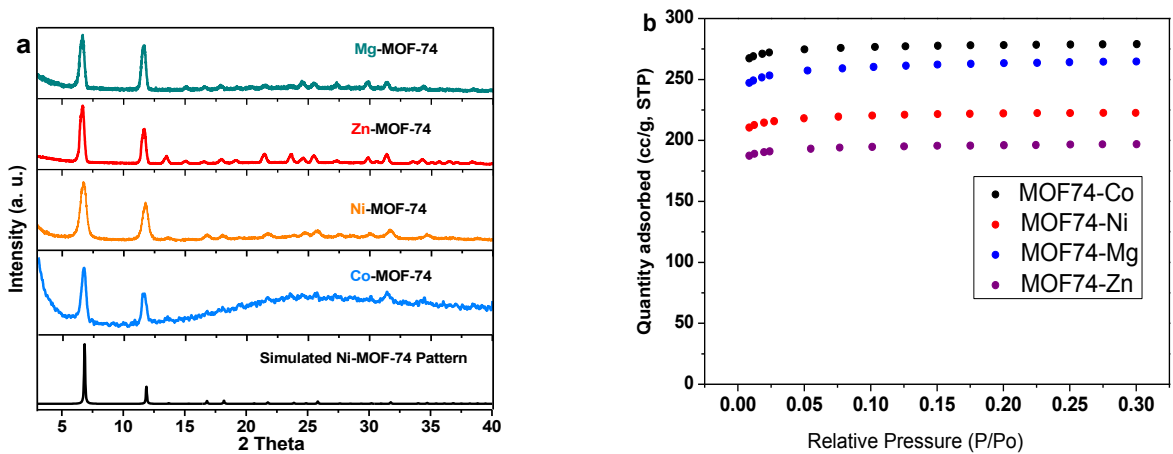

**Figure S1**. (a) Power X-ray diffraction pattern of MOF-74 samples (after solvent exchange) with the simulated pattern from single crystal data of Ni-MOF-74 taken from reference (1). (b) $N_2$ adsorption in M-MOF-74 (M= Mg, Zn, Co, Ni)

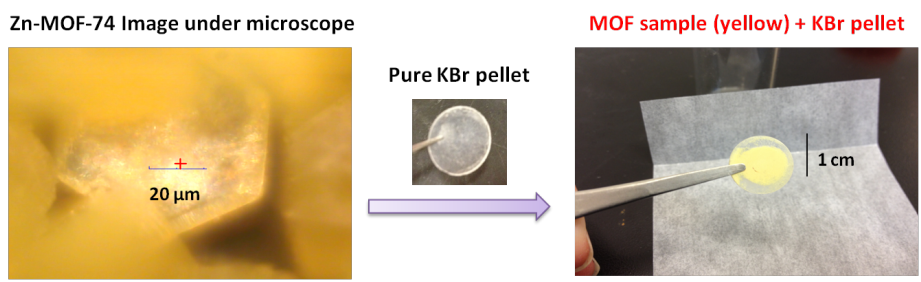

**Figure S2**. MOF powder crystals pressed onto the KBr pellet.

## 2. IR and Raman spectra of MOF phonon modes

To characterize the interaction of water molecules with framework structures, it is necessary to make an assignment of the vibrational modes. The detailed assignments are derived by comparing the spectra of MOF-74 (see Figures 7, S3 and S7) to those of free dihydroxyterephthalic acid molecules (see Figure S13) and referring to studies about hydroxybenzoic acid (salicylic acid) in the literature.[9-14] Coordination of ligand with metal ions brings about some characteristic changes in the IR spectra in comparison with the spectra of pure ligands—stretching modes of the carbonyl group $\nu(C=O)$ disappear, being replaced by other strong broad peaks around 1540 to 1580 cm⁻¹ and 1355 to 1371 cm⁻¹, which correspond to the asymmetric ($\nu_{as}$) and symmetric mode ($\nu_s$) of the carboxylate group $COO^-$.[9, 10, 14] Carboxlyate bending modes $\beta_{as}$ and $\beta_s$ appear in the lower frequency region 570 to 822 cm⁻¹.[9] This region was also characterized by strong CH out-of-plane deformation modes $\gamma(C-H)$.[9] The in-plane deformation modes $\beta(C-H)$ fall into a higher frequency region between 1119 to 1336 cm⁻¹. $\nu(C-O)$ in the phenolic group shifts to a higher frequency region around 1240 to 1250 cm⁻¹ upon depotonation during the MOFs synthesis.[12-14] The benzene ring C=C stretching modes are observed in the region from 1407 to 1484 cm$^{-1}$.$^{13, 14}$ The ring out of deformation mode φ(C=C) is located in the region 631 to 650 cm$^{-1}$.$^{9}$ The assignment of other bands occurring in the IR spectra of MOF-74 compounds is presented in Table S1. The assignments of the modes observed in Raman spectra are summarized in Table S2.$^{9}$

**Table S1. Assignment of absorption bands from infrared spectra of activated Zn, Mg, Ni, and Co-MOF-74 compounds.**

| Zn | Mg | Ni | Co | Assignment |
|---|---|---|---|---|
| 1552 | 1580 | 1544 | 1540 | ν$_{as}$(COO$^-$) |
|  | 1484 |  |  | ν(CC)ar |
| 1457 | 1449 | 1445 | 1454 | ν(CC)ar |
| 1425 | 1429 | 1407 | 1407 | ν(CC)ar |
| 1355 | 1371 | 1361 | 1363 | ν$_s$(COO$^-$) |
|  | 1336 |  |  | β(C-H) ip |
| 1246 | 1238 | 1239 | 1241 | ν(C-O)phenolate |
| 1197 | 1211 | 1196 | 1194 | β(C-H)ip |
| 1119 | 1123 | 1119 | 1122 | β(C-H) ip |
| 902 | 910 | 909 | 908 | γ(C-H) |
| 886 | 895 | 884 | 888 | γ(C-H) |
| 829 | 829 | 822 | 823 | γ(C-H) |
| 799 | 822 | 811 | 810 | β$_{as}$(COO-) |
| 650 | 632 | 639 | 631 | φ(C=C) |
| 570 | 587 | 590 | 587 | β$_s$(COO-) |

Note: "as" and "s" correspond to asymmetric and symmetric modes. "ar" refers to aromatic ring stretching mode.

**Table S2. Assignment of vibrational bands from Raman spectra of activated Zn, Mg, Ni, and Co-MOF-74 compounds.**

| Zn | Mg | Ni | Co | Assignment |
|---|---|---|---|---|
| 1620 | 1626 | 1619 | 1619 | ν$_{8a}$(CC)ar |
| 1565 | 1577 | 1564 | 1562 | ν$_{as}$(COO$^-$) |
| 1506 | 1507 | 1503 | 1506 | ν$_{19b}$(CC)ar |
| 1414 | 1429 | 1418 | 1415 | ν$_s$(COO$^-$) |
| 1273 | 1288 | 1279 | 1275 | ν(C-O)phenolate |
| 821 | 821 | 831 | 821 | β$_{as}$(COO-) |
| 565 | 564 | 574 | 566 | β$_s$(COO-) |

Note: 8a and 19b are Wilson numbering for benzene ring C=C stretching modes.

## 3. H₂O desorption from Mg-MOF-74 at 200 ˚C

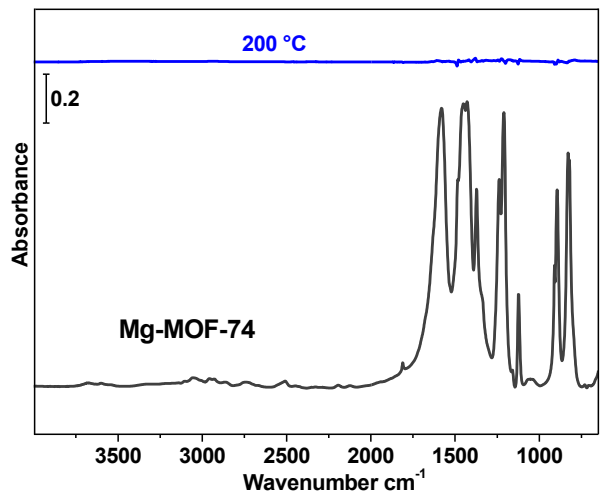

**Figure S3**. Difference spectra of hydrated Mg-MOF-74 after loading H$_2$O as a function of pressure up to 8 Torr at room temperature, evacuating the chamber to below 20 mTorr Torr for 100 min and then heating at 200 ˚C (as shown in Figure 1), compared to the original MOF spectra before loading water.

## 4. D₂O adsorption into Mg-MOF-74 at 24 ˚C and desorption upon annealing.

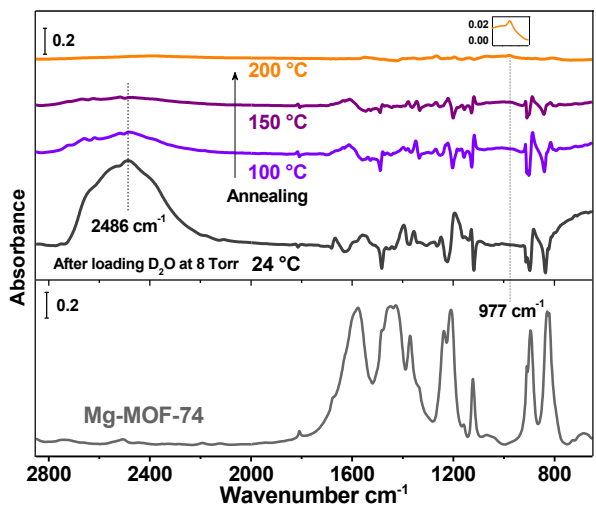

**Figure S4**. IR absorption spectra of D$_2$O adsorbed into Mg-MOF-74 at 8 Torr (black) and upon evacuation (20 mTorr base pressure) as a function of annealing temperature (100 ˚C, 150 ˚C and 200 ˚C). The inset shows the formation of weak 977 cm$^{-1}$ band during the evacuation process at 200 ˚C for 20 min.

      The D$_2$O was loaded into Mg-MOF-74 at room temperature by exposing the sample to 8 Torr vapor. The spectra shown in Figure S4 were taken after 10 min to make sure the adsorption reaches equilibrium. Similar to H$_2$O loading, the MOF phonon modes are significantly perturbed upon D$_2$O inclusion. The regeneration was performed by slowly heating the sample (6 ˚C/min) in vacuum (<20 mTorr base pressure). At 100 and 150 ˚C, the temperature was hold for 3 min; at 200 ˚C, the temperature

was kept for 20 min. Water molecules gradually desorb from the frameworks and the perturbations are also recovered upon water removal. The weak 977 cm$^{-1}$ band (due to the motion of deuterium atom) can be detected after annealing up to 200 ˚C, indicating the residual water molecules dissociate within MOFs. There is a competition between the departure (desorption) and reaction of water molecules within the frameworks during the annealing process. Most of molecularly adsorbed water desorbs from the frameworks below 150 ˚C (see Figure 1 and S4) when the annealing rate is low (<6 ˚C/min) because it allows any remaining water to desorb below 150 ˚C. Since the onset of dissociation reaction occurs at 150 ºC, it is still possible to detect some dissociation even at 6 ˚C/min, depending on the trace concentration of water.

## 5. H$_2$O and D$_2$O adsorption in MOF-74 (Zn, Ni, Co) at 100 ˚C, 150 ˚C and 200 ˚C

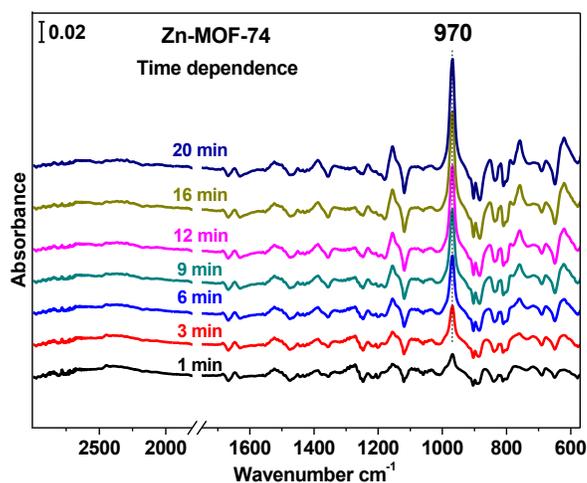

**Figure S5**. Time dependent IR absorption spectra of hydrated Zn-MOF-74 after introducing 8 Torr D$_2$O vapor within 20 min.

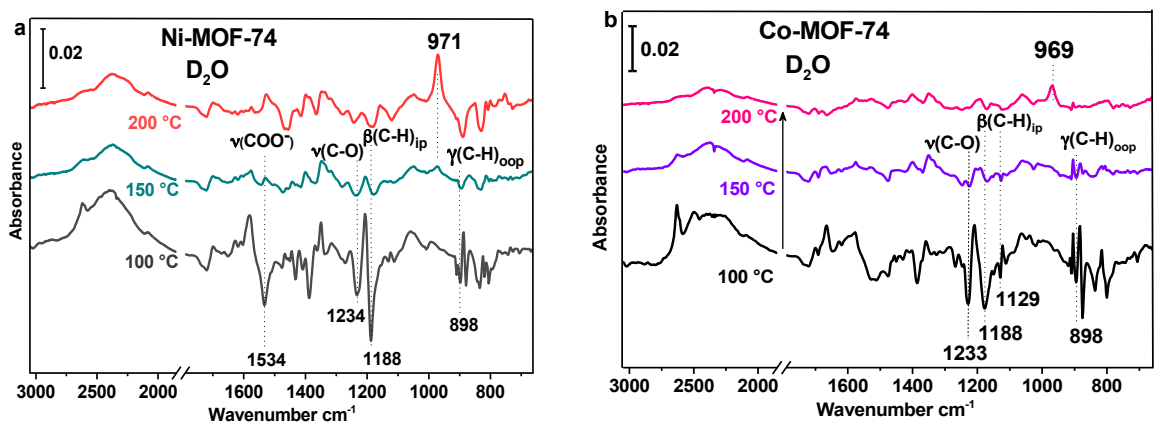

**Figure S6**. IR absorption spectra of Ni(a), Co(b)-MOF-74 hydrated by introduction of 8 Torr D$_2$O vapor for 20 min and evacuation of gas phase at 100 ˚C, 150 ˚C and 200 ˚C, referenced to the activated MOF in vacuum.

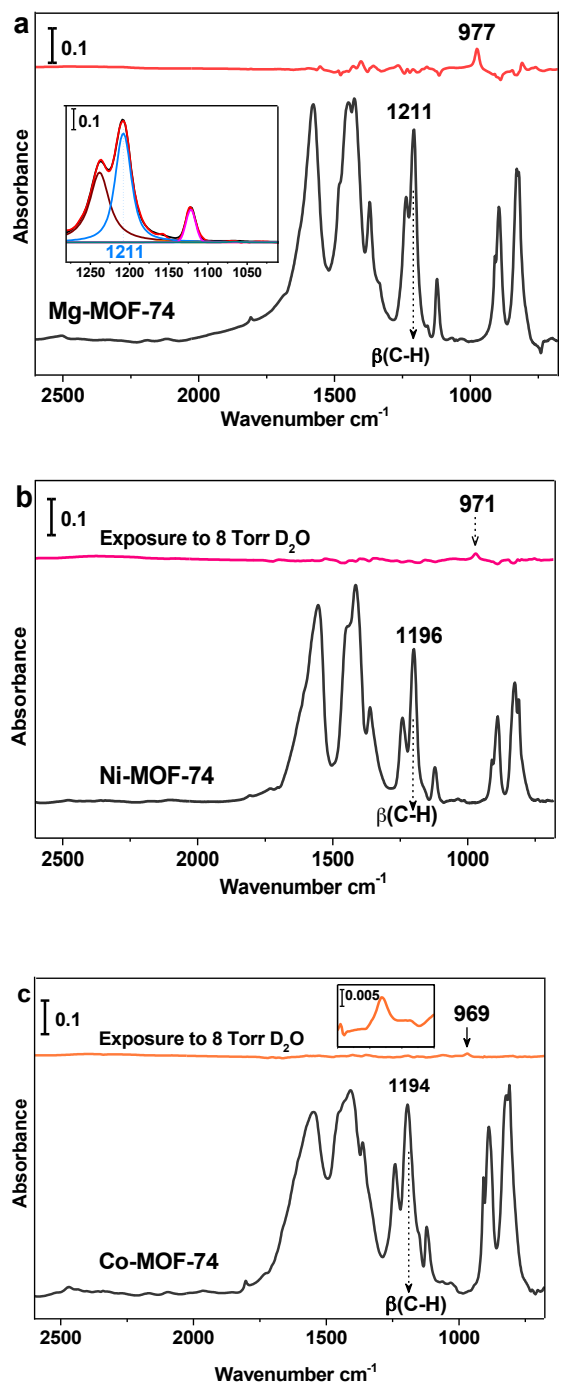

**Figure S7**. IR spectra of D$_2$O exposed Mg, Ni, Co-MOF-74 samples at 200 °C referenced to the corresponding clean activated MOFs, compared with the IR absorption spectra of the corresponding activated MOF-74 referenced to KBr pellets.

## 6. CO$_2$ adsorption in pristine and D$_2$O exposed samples.

Since the IR absorption of CO$_2$ gas at pressures above 10 Torr is prohibitively high (complete absorption saturation over the spectral range at and in the vicinity of the gas phase absorption), only up to 6 Torr CO$_2$ was introduced into the sample to measure the absorbance before and after D$_2$O exposure at 200 ˚C. For this pressure, the integrated areas of the CO$_2$ bands decrease by ~41 % of the initial value for Mg-MOF-74, ~23 % for Ni-MOF-74, and ~9% for Co-MOF-74 (see Figure S8).

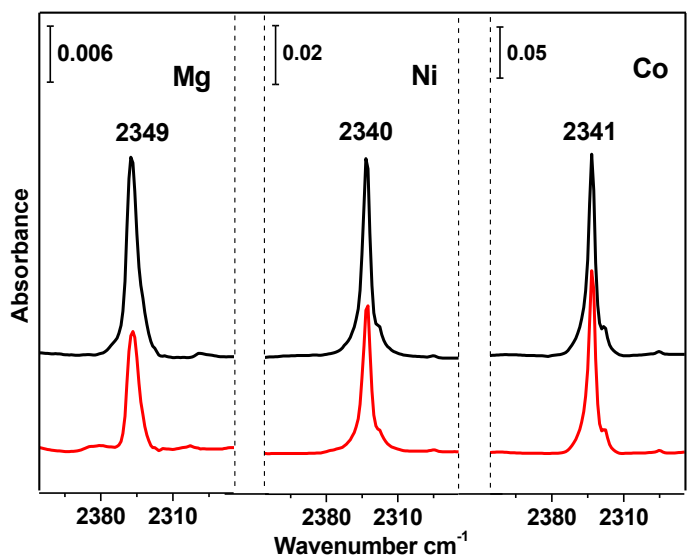

**Figure S8**. IR spectra of CO$_2$ adsorption into activated Mg, Ni, Co-MOF-74 (top, black) and D$_2$O exposed samples (bottom, red) at 200 ˚C for 20 min. All the measurements were taken after the samples were cooled back to room temperature.

## 7. Reproducibility of the measurements

The experiments were run twice on two different pellets prepared by the same method to check the reproducibility of the results. Figure S9 shows that relative intensities of 970 cm$^{-1}$ bands (after being normalized to β(CH) mode) in two different experiments are within 3 %.

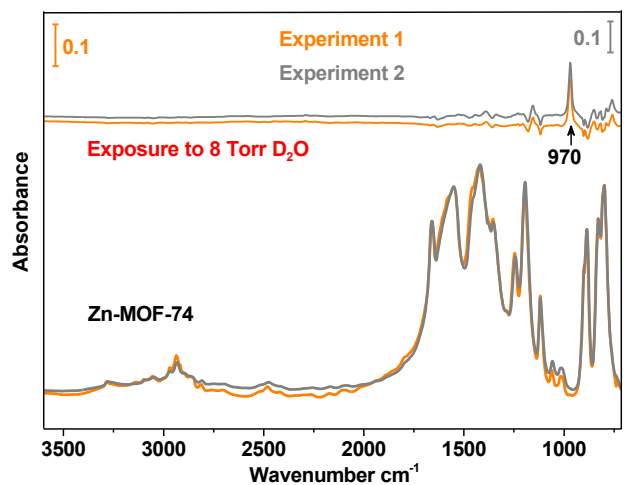

**Figure S9**. IR spectra of $D_2O$ exposed (20 min) Zn-MOF-74 at 200 °C referenced to the activated MOFs spectra (top), compared with the IR absorption spectra (bottom) of activated MOF-74 referenced to KBr pellet.

## 8. Zn-MOF-74 exposing to $D_2O$ after introducing 600 mtorr vapor (3%, RH).

After very low vapor pressure (600 mtorr, 3% RH) $D_2O$ exposure at 200 °C, the 970 cm$^{-1}$ band can be still detected with a normalized (refer to β(CH) band) intensity of 0.014 (see Figure S10). This indicates that Zn-MOF-74 has a high sensitivity to moisture at high temperatures. After $D_2O$ exposure, the $CO_2$ adsorption decreased by ~13% compared to the pristine MOFs.

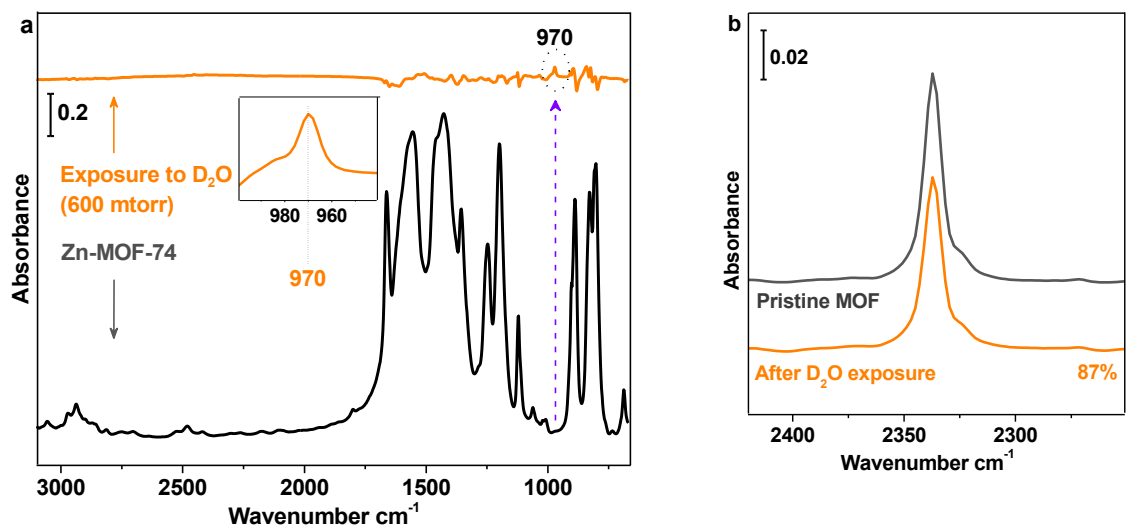

**Figure S10**, (a) IR spectra of 600 mtorr $D_2O$ (red) exposed Zn-MOF-74 at 200 °C referenced to the activated MOFs spectra, compared with the IR absorption spectra of activated MOF-74 (black) referenced to KBr pellet. (b) IR spectra of $CO_2$ adsorption into activated Zn-MOF-74 (pristine, top) and 600 mtorr $D_2O$ (bottom) exposed sample.

## 9. Effect of multiple $D_2O$ exposures on $CO_2$ adsorption in Mg-MOF-74

$CO_2$ adsorption in Mg-MOF-74 was further investigated by repeated $D_2O$ exposures. For each cycle, the Mg-MOF-74 was exposed to a 9.3 Torr $D_2O$ vapor for 20 min at 200 °C and the chamber was then evacuated. The IR absorption measurement was taken at 6 Torr after the sample cooled back to room temperature (24 °C) to determine the $CO_2$ concentration. In the first cycle, the $CO_2$ absorbance decreased by ~44 % of the initial value as shown in Figure S11. This figure shows that after four cycles, the $CO_2$ adsorption measured by IR continually diminished to ~39 % of the pristine sample. The $CO_2$ frequency red shifts slightly by 2 cm$^{-1}$ in the later cycles, this could be due to the changed chemical environment after the $D_2O$ reaction within the framework. Figure S11b shows that the integrated areas of the 977 cm$^{-1}$ band increases during the cyclic exposure, while $CO_2$ absorption band decreases.

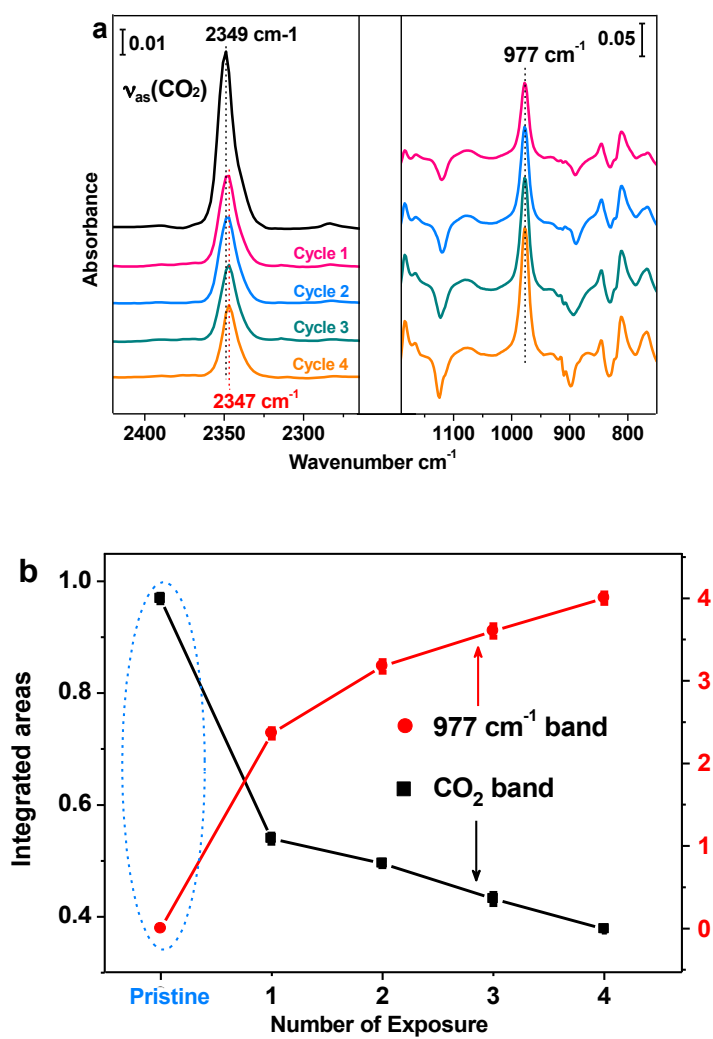

**Figure S11**. (a) IR spectra of $CO_2$ adsorption in Mg-MOF-74 at 6 Torr and the band at 977 cm$^{-1}$ after cyclic $D_2O$ exposure. (b) Evolution of the integrated areas of the adsorbed $CO_2$ band at 2349 and the deuterium band at 977 cm$^{-1}$ in Mg-MOF-74.

## 10. Raman spectra of MOF-74 before and after D₂O exposure

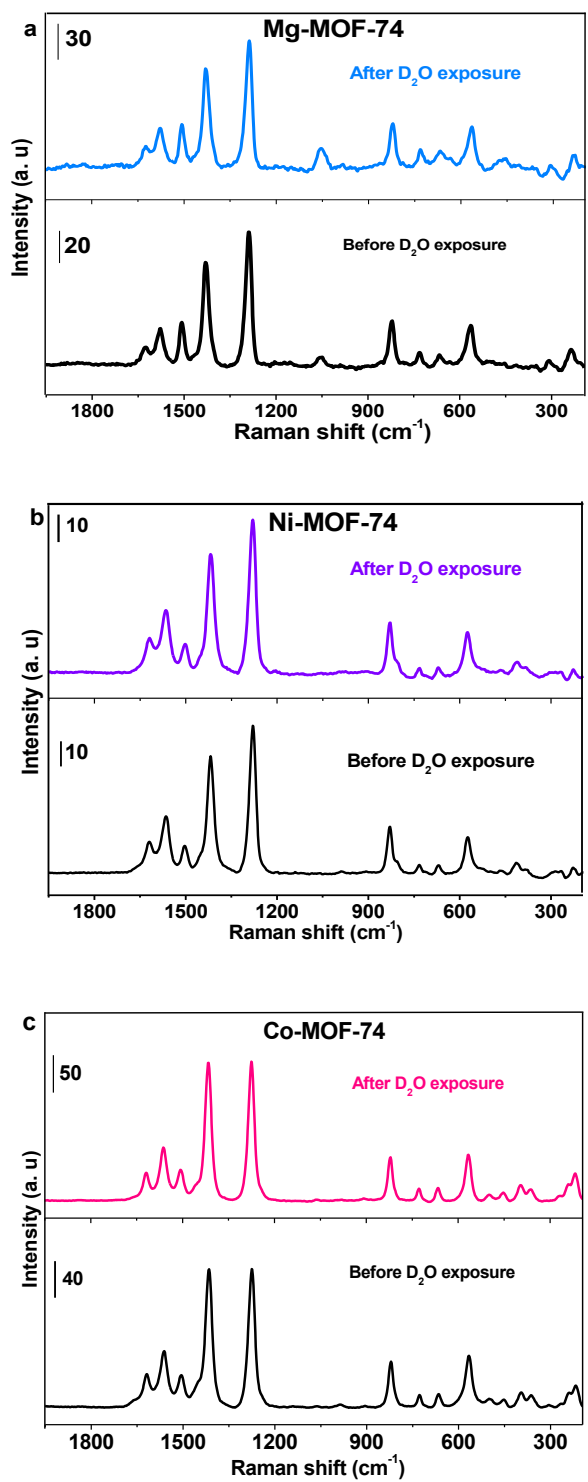

**Figure S12**. Raman spectra of as-synthesized MOF-74 (Mg, Ni, Co) samples and hydrated MOF materials after exposing to 8 Torr D₂O vapors at 200 °C. The spectra were recorded at room temperature.

## 11. D₂O exposure to pure dobdc ligand at 200 °C

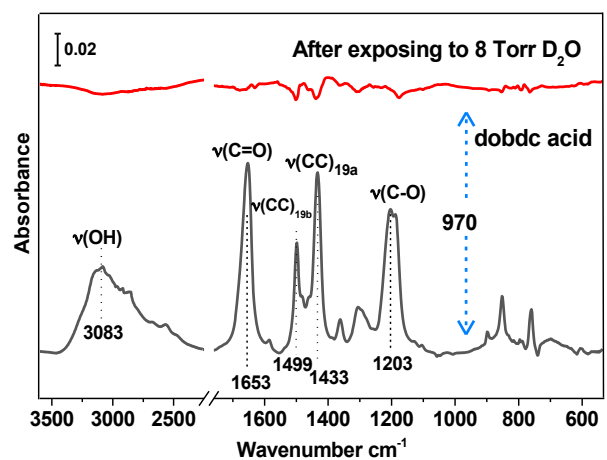

**Figure S13**. IR spectra of D₂O exposed dobdc (dobdc4− = 2,5-dioxido-1,4-benzenedicarboxylate) ligand referenced to the activated powder, compared with the IR absorption spectra of activated powder dobdc ligand referenced to KBr pellet.

## 12. Stability of OD group upon exposure to water ($H_2O$) vapor

By the $H_2O$ exposure experiment on the D-exchanged sample (Zn-MOF-74), we have shown (Figure S14) that at room temperature, the band at 970 cm$^{-1}$ is perturbed (shift slightly to a lower frequency) upon adsorption of $H_2O$ molecules at 8 Torr. At 200 °C, after exposure to $H_2O$ for 20 min, the band decreased a little, suggesting that only a minority of D atoms are exchanged with $H_2O$.

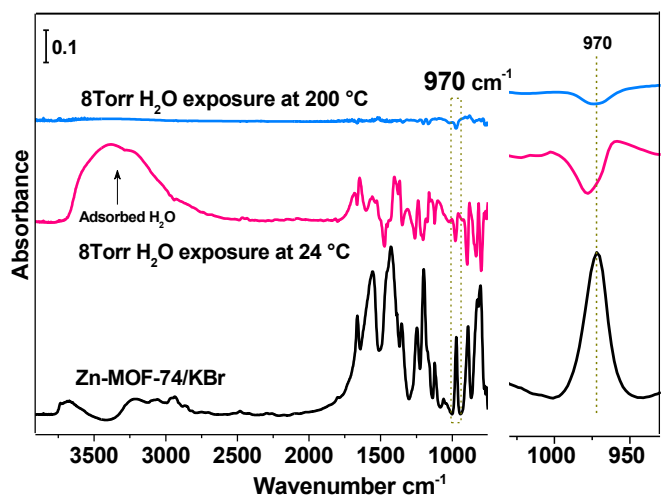

**Figure S14.** D exchanged Zn-MOF-74 spectra referenced to activated MOF in vacuum. Middle: After exposing D exchanged Zn-MOF-74 to 8 Torr $H_2O$ at 24 °C for 50 min. Top: After exposing D exchanged Zn-MOF-74 to 8Torr $H_2O$ at 200 °C for 50 min.

## 13. Calculated vibrational modes

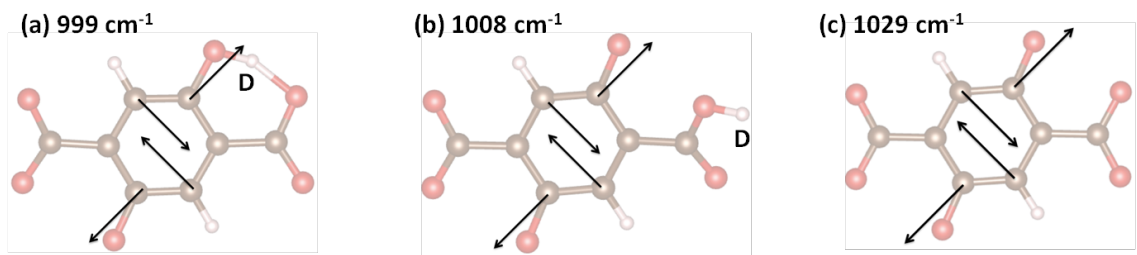

**Figure S15**. Calculated vibrational modes for (a) adsorption site 1 at 999 cm$^{-1}$; (b) adsorption site 2 at 1008 cm$^{-1}$; and (c) pristine MOFs at 1029 cm$^{-1}$.

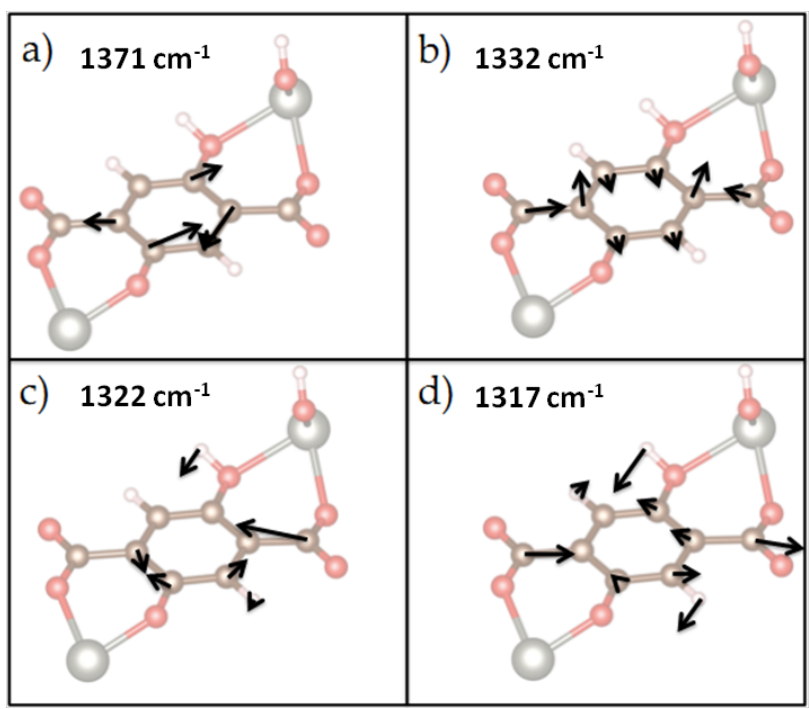

**Figure S16**. Calculated vibrational modes for "linker + D" at (a) 1371 cm$^{-1}$, (b) 1332 cm$^{-1}$, (c) 1322 cm$^{-1}$, (d) 1317 cm$^{-1}$. The black arrows represent the eigenvector of the vibrational mode.